\documentclass[%
reprint,
 amsmath,
 amssymb,
 aps,
 pra,
twocolumn,
]{revtex4-1} 

\usepackage{graphicx}
\usepackage{amsmath}
\usepackage{amsfonts}
\usepackage{dcolumn}
\usepackage{siunitx}
\DeclareSIUnit{\sqrthz}{\ensuremath{\sqrt{\text{\hertz}}}}
\usepackage{multirow}
\usepackage{hyperref}

\hypersetup{
    colorlinks=true,       
    linkcolor=blue,          
    citecolor=blue,        
    filecolor=blue,      
    urlcolor=blue           
}

\graphicspath{{./}}

\usepackage{txfonts}

\begin{document}

\preprint{APS/123-QED}

\title{Light-induced atomic desorption of lithium}

\author{D. S. Barker}
\email[]{daniel.barker@nist.gov}
\affiliation{Sensor Science Division, National Institute of Standards and Technology, Gaithersburg, MD 20899, USA}
\author{E. B. Norrgard}
\author{J. Scherschligt}%
\author{J. A. Fedchak}
\author{S. Eckel}
\email[]{stephen.eckel@nist.gov}
\affiliation{Sensor Science Division, National Institute of Standards and Technology, Gaithersburg, MD 20899, USA}

\date{\today}

\begin{abstract}

We demonstrate loading of a Li magneto-optical trap using light-induced atomic desorption.
The magneto-optical trap confines up to approximately \(4\times10^{4}\)~\(^{7}\text{Li}\)~atoms with loading rates up to approximately \(4\times10^{3}\) atoms per second.
We study the Li desorption rate as a function of the desorption wavelength and power.
The extracted wavelength threshold for desorption of Li from fused silica is approximately \(470~\si{\nano\metre}\).
In addition to desorption of lithium, we observe light-induced desorption of background gas molecules.
The vacuum pressure increase due to the desorbed background molecules is \(\lesssim50~\si{\percent}\) and the vacuum pressure decreases back to its base value with characteristic timescales on the order of seconds when we extinguish the desorption light.
By examining both the loading and decay curves of the magneto-optical trap, we are able to disentangle the trap decay rates due to background gases and desorbed lithium.
Our results show that light-induced atomic desorption can be a viable Li vapor source for compact devices and sensors.

\end{abstract}

\maketitle

\section{\label{Sec:Intro}Introduction}

There is a growing interest in miniaturization of laser-cooled atomic technologies~\cite{Rushton2014, Keil2016}.
Laser-cooled atoms are a promising candidate for the realization of a variety of portable devices, including quantum repeaters~\cite{Duan2001, Yang2016a}, atom interferometers~\cite{Bodart2010, Barrett2016, Abend2016, Wu2017}, and vacuum sensors~\cite{Arpornthip2012a, Scherschligt2017}.
Most work toward mobile laser-cooled atom devices has focused on ultracold Rb and Cs due to their high room-temperature vapor pressure and the proliferation of laser technology at the necessary wavelengths~\cite{Bodart2010, Barrett2016, Abend2016, Schkolnik2016, Wu2017}.
However, devices based on laser-cooled Li would be advantageous for vacuum sensing~\cite{Scherschligt2017}, due to lithium's low room-temperature vapor pressure (\(\lesssim10^{-17}~\si{\pascal}\)~\cite{Alcock1984}).

A central challenge to the miniaturization of laser-cooled atomic devices is outgassing of the atomic vapor source~\cite{Rushton2014}.
Mobile sensors will likely have minimal vacuum pumping, so excessive outgassing of any \textit{in-vacuo} component will limit the useful lifetime of the device.
In applications such as vacuum sensing, where a cold-atom vacuum sensor would be attached to a larger vacuum system, atomic source outgassing could easily limit sensor performance.
Lithium's vapor pressure at room temperature precludes the use of a vapor cell~\cite{Anderson2001}.
Lithium dispenser sources must operate at high temperatures to produce appreciable vapor pressure; leading to high outgassing rates and limiting the achievable vacuum pressure~\cite{Fortagh1998, Norrgard2018}.
An attractive alternative to these conventional vapor sources is light-induced atomic desorption (LIAD)~\cite{Abramova1984, Gozzini1993}, where atoms are liberated from a surface using photons.
In the context of laser-cooled atomic gases, the desorption surface is typically a glass or metal vacuum chamber wall and the desorption light is usually generated by a short wavelength incoherent source~\cite{Anderson2001, Klempt2006, Zhang2009, Coppolaro2014}.
The extra gas load from the desorption process is rapidly reduced when the desorption light source is extinguished.
Among elements amenable to laser cooling, LIAD has previously been observed for calcium~\cite{Mango2008} and all alkali metals except Li~\cite{Abramova1984, Gozzini1993, Meucci1994, Alexandrov2002, Gozzini2004, Coppolaro2014}.
It can efficiently load magneto-optical traps (MOTs) with high atom number~\cite{Anderson2001, Atutov2003, Aubin2005, Klempt2006, Telles2010, Torralbo-Campo2015, Agustsson2017}, allowing for the production of quantum degenerate gases~\cite{Du2004, Mimoun2010}.

We report and detail the first observed light-induced atomic desorption of lithium.
We characterize the desorption of \(^{7}\text{Li}\) atoms from a fused silica window for three light sources with distinct operating wavelengths.
The desorbed atoms are captured in a six-beam magneto-optical trap.
The largest MOT loading rate, approximately \(4\times10^{3}\) atoms per second, occurs when a \(385~\si{\nano\metre}\) light-emitting diode (LED) induces atomic desorption.
At this loading rate the MOT population reaches approximately \(4\times10^{4}\)~\(^{7}\text{Li}\)~atoms.
Light induced atomic desorption is often explained by analogy to the photoelectric effect and prior studies have found a quadratic dependence of the desorption yield on source wavelength~\cite{Fowler1931, DuBridge1933, Xu1996, Gozzini2004, Klempt2006}.
Our data are consistent with a quadratic dependence of the MOT loading rate on LIAD wavelength, from which we infer the threshold wavelength for LIAD of \(^{7}\text{Li}\).
Our results show that LIAD is a viable atom source for lithium-based vacuum sensors and may be useful for other compact devices, depending on the atom number requirements.

We describe the measurement apparatus in Section~\ref{Sec:Apparatus}.
Section~\ref{Sec:Results} shows the experimental data and contains a discussion of the results.
We have studied the variation in the MOT loading rate and atom number as a function of the LIAD source power and wavelength.
We summarize our findings and discuss future outlook in Section~\ref{Sec:Discussion}.

\section{\label{Sec:Apparatus}Apparatus}

We characterize light-induced atomic desorption process by loading a Li MOT within a stainless-steel vacuum chamber.
All steel components of the chamber were vacuum baked at \(425~\si{\degreeCelsius}\) for 21 days to reduce hydrogen outgassing~\cite{Sefa2017, Fedchak2018}.
We did not bake the chamber after assembly to remove \(\text{H}_{2}\text{O}\).
However, the chamber was held under vacuum for several months before LIAD studies began, so its outgassing rate and base pressure are similar to those that would have been achieved after a \(48~\si{\hour}\) bake at \(150~\si{\degreeCelsius}\).
A \(50~\si[per-mode = symbol]{\liter\per\second}\) ion pump removes background gases from the vacuum chamber.
The base vacuum pressure, as measured by a metal-envelope enclosed Bayard-Alpert ionization gauge~\cite{Arnold1994, Fedchak2012}, is \(4(1)\times10^{-8}~\si{\pascal}\) (Here, and throughout this paper, parenthetical quantities represent standard deviations).
An alkali metal dispenser made from 3D-printed titanium~\cite{Norrgard2018} was used to deposit lithium on the vacuum chamber's fused silica viewports.
Immediately after deposition, the optical depth of each viewport's lithium coating was on the order of \(0.1\).
We observed no reduction in the coating's optical depth during our LIAD study.
The viewports are not anti-reflection coated for the MOT's operating wavelength to allow for better transmission of LIAD light.

The MOT operates on the \(^2\)S\(_{1/2}\) \((F=2)\) to \(^2\)P\(_{3/2}\) \((F=3)\) transition of the \(^{7}\text{Li}\).
It comprises six independent, circularly polarized, laser beams and a set of N52-grade neodymium-iron-boron magnets.
Each laser beam has \(40(1)~\si{\milli\watt}\) of power, a Gaussian \(1/e^{2}\) radius of \(7.1(4)~\si{\milli\metre}\), and a detuning of \(-18~\si{\mega\hertz}\) from the \(F=2\rightarrow F'=3\) transition.
Two 3D-printed thermoplastic mounts secure the permanent magnets to the vacuum chamber such that they produce a quadrupole magnetic field with a vertical gradient of \(3~\si[per-mode = symbol]{\milli\tesla\per\centi\metre}\).
An electro-optic modulator (EOM) provides repumping for the MOT by adding \(814~\si{\mega\hertz}\) radiofrequency sidebands to the MOT beams.
The \(+1\) order sideband addresses the \(F=1\rightarrow F'=2\) repump transition and contains approximately \(20~\si{\percent}\) of the optical power (The ratio of the carrier power to the \(+1\) order sideband power is approximately \(3{:}1\)).

We use three different light sources to desorb Li from the fused silica viewports.
The first two sources are multimode laser diodes (LD) operating at \(405~\si{\nano\metre}\) and \(445~\si{\nano\metre}\).
The \(405~\si{\nano\metre}\) LD has a fiber pigtail and can deliver up to \(350~\si{\milli\watt}\) to the vacuum chamber.
The \(445~\si{\nano\metre}\) LD is free-space coupled and has a maximum power output of \(1.6~\si{\watt}\).
Our last light source is a UV LED with a center wavelength of \(385~\si{\nano\metre}\) and a maximum power output of \(1.6~\si{\watt}\).
We collimate the LED output using an aspheric condenser lens, but the LED output's large divergence still limits the power available for LIAD to approximately \(500~\si{\milli\watt}\).
The average intensity corresponding to the maximum LIAD power is approximately \(45~\si[per-mode = symbol]{\milli\watt\per\square\centi\metre}\), \(70~\si[per-mode = symbol]{\milli\watt\per\square\centi\metre}\), and \(300~\si[per-mode = symbol]{\milli\watt\per\square\centi\metre}\) for the \(385~\si{\nano\metre}\), \(405~\si{\nano\metre}\), and \(445~\si{\nano\metre}\) light sources, respectively. 

A longpass dichroic mirror overlaps the LIAD light with one of the MOT laser beams to couple it onto the vacuum viewports.
The dichroic mirror has a cutoff wavelength of \(380~\si{\nano\metre}\), which prevents us from investigating desorption at shorter wavelengths.
All our LIAD light sources are collimated and normally incident to both the input and output viewports.
As such, LIAD light passes through the vacuum chamber without directly impinging on any stainless steel surfaces.
Fresnel and diffuse reflections from the Li-coated viewports lead to a small amount of desorption light eventually striking the interior of the vacuum chamber.
Limited tests of direct desorption from the vacuum chamber interior using the \(405~\si{\nano\metre}\) LD suggest that Li desorption from stainless steel is no better than Li desorption from fused silica.
Moreover, increased desorption of background gas molecules from the stainless steel resulted in significant reductions in the Li MOT population.
Because the illumination of stainless steel surfaces is substantially dimmer than the illumination of the viewports, we believe that it contributes minimally to the LIAD yield.

\section{\label{Sec:Results}Results}

We load our Li MOT using LIAD for \(40~\si{\second}\) while measuring the MOT fluorescence with a charge-coupled device (CCD) camera.
The long loading time ensures that the MOT population saturates to a final number \(N_{S}\) for all of the LIAD wavelengths and powers investigated here~\footnote{For optimal loading conditions, the MOT atom number reaches \(10^{4}\) atoms in about \(3~\si{\second}\) and saturates within \(20~\si{\second}\).}.
In the abscence of the LIAD illumination, there is no observable MOT.\@
After the MOT loads completely, we extinguish the LIAD light and record the decay of the trapped atom number for \(10~\si{\second}\).
We activate the LIAD light source and MOT beams \(5~\si{\second}\) before turning on the repump EOM to collect images for background subtraction.
Our estimated non-statistical uncertainty in the conversion from integrated CCD counts to atom number is \(50~\si{\percent}\).

\begin{figure}
\includegraphics[width=\linewidth]{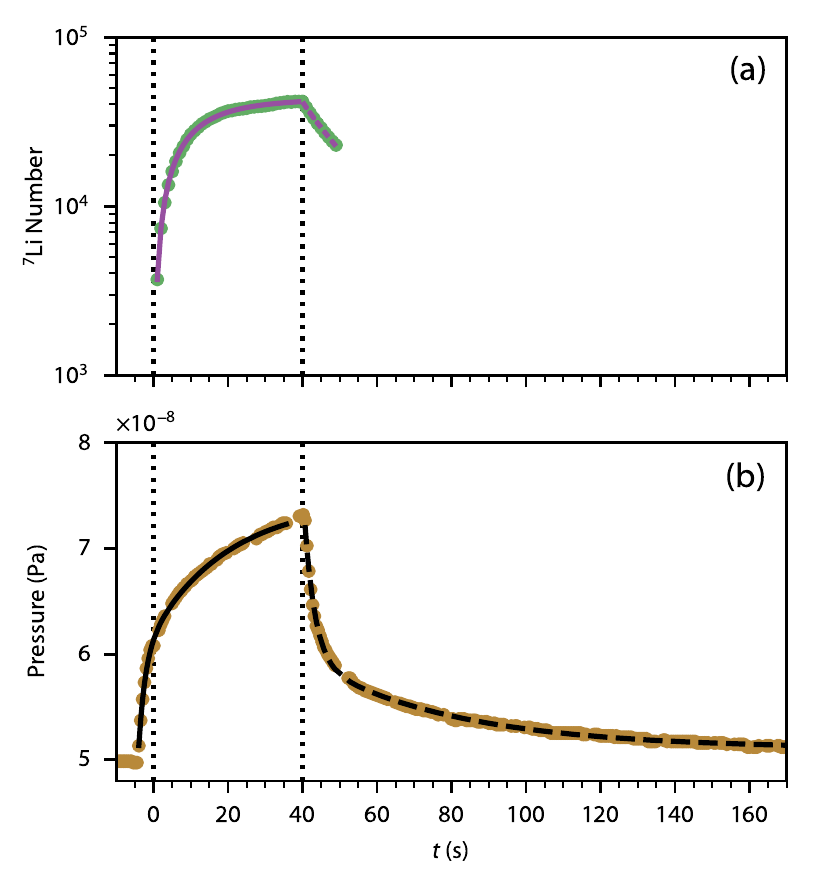}
\caption{MOT loading with \(490~\si{\milli\watt}\) of light from the \(385~\si{\nano\metre}\) LED.\@
(a) shows the atom number as a function of time in green and fits to the data using Equations~(\ref{MOTload}) and~(\ref{MOTdecay}) in purple, with solid lines indicating loading and dashed lines indicating decay.
The vacuum pressure measured by an ionization gauge is shown in (b).
The solid (dashed) black line in the bottom subplot indicates a double exponential growth (decay) fit to the pressure data (see Equations~(\ref{Prise}) and~(\ref{Pdecay})).
The vertical dotted lines denote the beginning and end of MOT loading.\label{LIADrun}}
\end{figure}

Figure~\ref{LIADrun} (a) shows a typical MOT loading curve taken using the UV LED to desorb lithium atoms.
The corresponding vacuum pressure dynamics (see Fig.~\ref{LIADrun} (b)) measured by the ionization gauge are not caused by desorbed Li.
The geometry of the vacuum chamber and lithium's low room-temperature vapor pressure prevent any lithium from reaching the ion gauge: the optimal path from the viewport to the gauge still requires five collisions with the walls of the vacuum chamber.
The sticking coefficient of Li on stainless steel and Li is \(1\) to an excellent approximation~\cite{Sugai1995, Kleine1995, Farmer2009, Skinner2016, Knaster2016}, so we conservatively estimate that the probability for a lithium atom to reach the ion gauge is \(\lesssim 10^{-10}\).
Additionally, when lithium was initially deposited onto the viewport from the dispenser, a turbomolecular pump and a residual gas analyzer (RGA) were attached to the vacuum chamber near the ion gauge.
The RGA detected no lithium despite the lithium vapor pressure, as measured by the MOT loading rate, being orders of magnitude higher than our LIAD setup can produce~\cite{Norrgard2018}.

Light-induced desorption of vacuum contaminants from the viewports causes the observed pressure variation~\cite{Halama1991, Herbeaux1999, Koebley2012}.
The pressure rise (fall) is well-described by a double exponential growth (decay), which has previously been observed to be characteristic of LIAD~\cite{Anderson2001, Klempt2006}.
We fit the pressure growth, \(P_{r}(t)\), and decay, \(P_{d}(t)\), using
\begin{equation}
P_{r}(t) = P_{f}(1-e^{-t/\tau_{f}})+P_{s}(1-e^{-t/\tau_{s}})+P_{b},
\label{Prise}
\end{equation}
and
\begin{equation}
P_{d}(t) = P_{1}e^{-t/\tau_{f}}+P_{2}e^{-t/\tau_{s}}+P_{b}.
\label{Pdecay}
\end{equation}
Here, \(P_b\) is the nominal pressure without LIAD;\@ \(P_{f}\) and \(P_{s}\) are the asymptotic pressure increases with characteristic timescales \(\tau_{f}\) and \(\tau_{s}\), respectively; and \(P_{1(2)} = P_{f(s)}(1-e^{-t_{\rm load}/\tau_{f(s)}})\) are the measured pressures after the MOT has loaded for \(t_{\rm load} = 40~\si{\second}\).
We find that \(\tau_{s}=30(3)~\si{\second}\) and \(\tau_{f}=3.2(2)~\si{\second}\) are independent of the LIAD wavelength and power.
The fast (slow) timescale, \(\tau_{f}\) (\(\tau_{s}\)), has been associated with adsorption of the LIAD product back onto vacuum chamber surfaces that have less than (more than) a monolayer coating of the LIAD product~\cite{Klempt2006}.
This explanation is reasonable provided that the adsorption time constants are smaller than the vacuum pumping time constant, \(\tau_{\rm pump}\), which is given by the volume of the vacuum system divided by the effective pumping speed~\cite{Jitschin2016}.
Because adsorption and vacuum pumping act in parallel, the LIAD pressure dynamics should occur on a timescale faster than the pumping time constant.
The \(\text{N}_{2}\) pumping speed of our ion pump and the volume of our vacuum system imply that \(\tau_{\rm pump}\approx100~\si{\milli\second}\).
The discrepancy between the measured and expected timescales hints that either the dominate non-Li desorption product is inefficiently removed by the ion pump or that some desorption continues even after the desorption light is switched off.

The MOT atom number, \(N\), increases during loading according to the differential equation
\begin{equation}
\frac{dN}{dt}=R-(KP_{r}(t-t_{0})+\Gamma_{\text{Li}})N,
\label{MOTload}
\end{equation}
where \(R\) is the MOT loading rate, \(t_{0}=-5~\si{\second}\) is the delay between LIAD source activation and the beginning of MOT loading, and \(K\) is a constant that relates the measured background pressure to the MOT loss rate.
We assume that the additional MOT loss due to desorbed lithium, with rate \(\Gamma_{\text{Li}}\), is time independent.
Because the MOT immediately begins to decay once the LIAD light is removed and the decay does not accelerate (which would imply that lithium vapor remained in the vacuum chamber), the lithium vapor pressure must decay on a timescale \(\lesssim1~\si{\second}\).
The rise and fall times for the lithium vapor pressure are identical since the effective vacuum pumping speed is independent of the LIAD process, so \(\Gamma_{\text{Li}}\) will reach a steady-state value before we activate the MOT.\@
Due to the chamber geometry and lithium's high sticking coefficient (see above), all Li pumping in our vacuum system is provided by the chamber surfaces close to the LIAD source and MOT.\@
When we extinguish the LIAD source, the MOT atom number decays as
\begin{equation}
\frac{dN}{dt}=-KP_{d}(t)N.
\label{diffDecay}
\end{equation}
The solution to Eq.~(\ref{diffDecay}) is
\begin{equation}
N(t)=N_{0}e^{-K\big(P_{1}\tau_{f}(1-e^{-t/\tau_{f}})+P_{2}\tau_{s}(1-e^{-t/\tau_{s}})+P_{bg}t\big)},
\label{MOTdecay}
\end{equation}
where \(N_{0}\) is the atom number the LIAD source is turned off.
By fitting each MOT decay curve with Eq.~(\ref{MOTdecay}) using parameters from the pressure decay (see Eq.~(\ref{Pdecay}) and Fig.~\ref{LIADrun} (b)), we can extract the proportionality constant, \(K\).
Using this value for \(K\) and the best fit parameters in \(P_{r}(t-t_{0})\), Eq.~(\ref{MOTload}) is fit to the associated MOT loading curve via numerical integration.
The fit yields the loading rate \(R\) and lithium-induced decay rate \(\Gamma_{\text{Li}}\) for each loading curve.

\begin{figure}
\includegraphics[width=\linewidth]{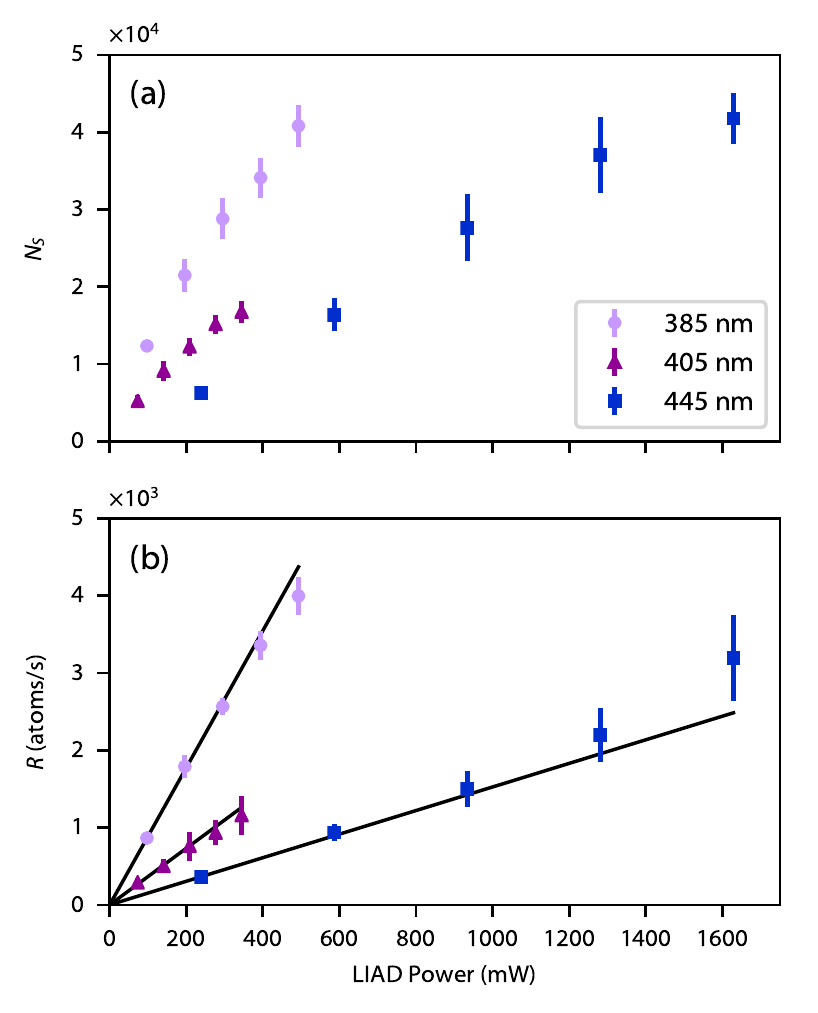}
\caption{Saturated atom number (a) and loading rate (b) for the \(^{7}\text{Li}\) MOT as a function of LIAD power.
Data are for LIAD sources operating at \(385~\si{\nano\metre}\) (lavender circles), \(405~\si{\nano\metre}\) (purple triangles), and \(445~\si{\nano\metre}\) (blue squares).
The solid lines in (b) are linear fits to the measured loading rate.
The errorbars in both subplots represent the standard deviation of at least four measurements.\label{LIADpower}}
\end{figure}

We measured the saturated atom number, \(N_{S}=R/(KP_{r}(t_{\rm load}-t_{0})+\Gamma_{\text{Li}})\), and loading rate, \(R\), of the MOT as function of the LIAD power and wavelength.
Figure~\ref{LIADpower} shows the results of these measurements.
The MOT can load as many as approximately \(4\times10^{4}\)~\(^{7}\text{Li}\)~atoms using the \(385~\si{\nano\metre}\) LED or \(445~\si{\nano\metre}\) LD as the LIAD light source.
Inducing desorption with the \(385~\si{\nano\metre}\) LED yields the fastest loading rates (up to approximately \(4\times10^{3}\) atoms per second).
Prior studies of LIAD of other alkali elements have reported saturation of both \(N_{S}\) and \(R\)~\cite{Klempt2006, Zhang2009}.
We do not observe saturation of either the MOT loading rate or the saturated atom number, except possibly when desorbing Li with the \(405~\si{\nano\metre}\) and \(445~\si{\nano\metre}\) LDs.
The lack of saturation implies that our LIAD light sources are not depleting the viewport's lithium coating~\cite{Torralbo-Campo2015}, which agrees with our observation that the viewports' optical depth is constant (see Section~\ref{Sec:Apparatus}).
Linear fits to the loading rate data yield slopes of \{\(8.9\), \(3.7\), \(1.5\)\}~\(\si{\per\milli\watt\per\second}\) for the \{\(385~\si{\nano\metre}\), \(405~\si{\nano\metre}\), \(445~\si{\nano\metre}\)\} light source. 
A parabolic fit to the loading rate slope as a function of LIAD photon energy suggests that there is a threshold wavelength for Li LIAD near \(470~\si{\nano\metre}\)~\cite{Xu1996, Gozzini2004, Klempt2006}.
The quadratic dependence of the loading rate on wavelength is justified by analogy to the photoelectric effect, but there is disagreement on the validity of this analogy in the literature~\cite{Xu1996, Gozzini2004, Marinelli2006, Lucchesini2016}.
However, the majority of experiments support both the existence of a threshold wavelength and the quadratic dependence of desorption rate on photon energy.

\begin{figure}
\includegraphics[width=\linewidth]{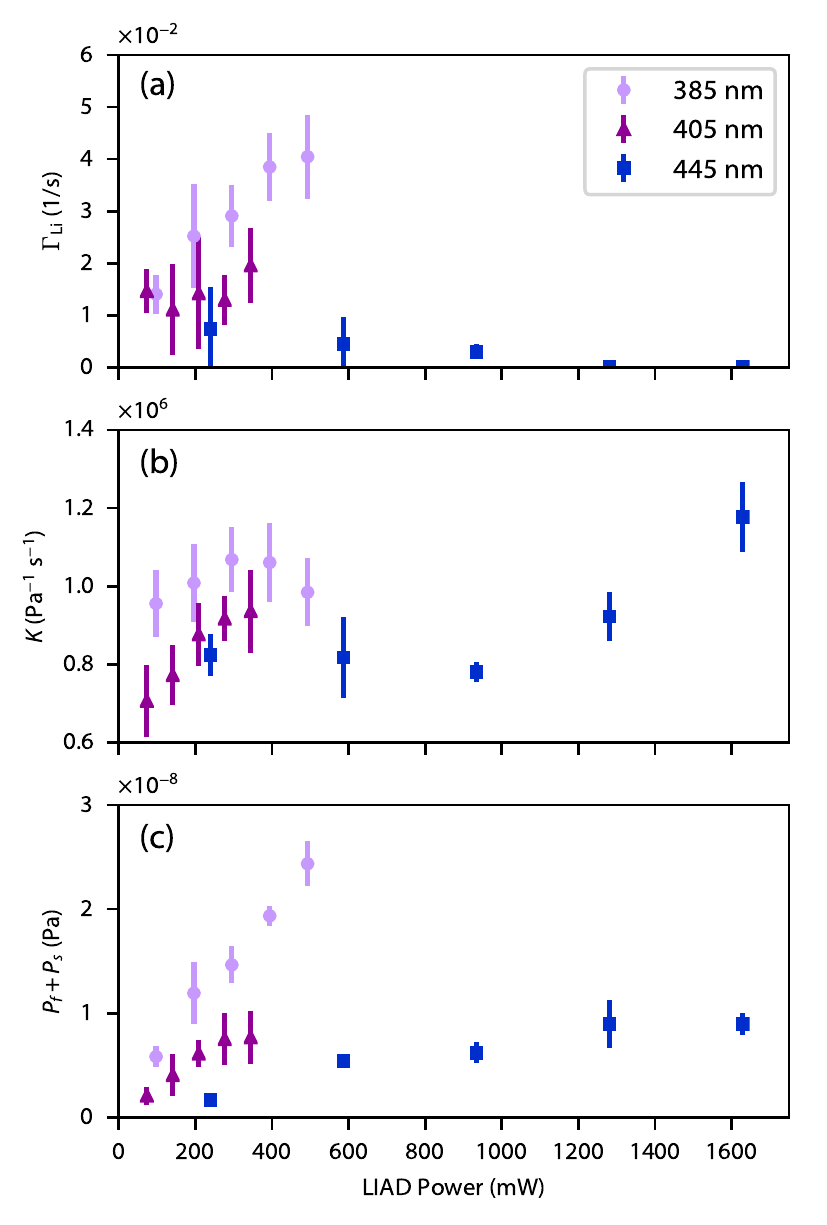}
\caption{Loss processes and pressure rise during LIAD as a function of power: (a) Loss rate, \(\Gamma_{\text{Li}}\), of \(^{7}\text{Li}\) from the MOT due to untrapped \(^{7}\text{Li}\), (b) background loss rate coefficient \(K\), and (c) total asymptotic pressure increase \(P_{f}+P_{s}\).
Data are for LIAD sources operating at \(385~\si{\nano\metre}\) (lavender circles), \(405~\si{\nano\metre}\) (purple triangles), and \(445~\si{\nano\metre}\) (blue squares).
The errorbars in each subplot represent the standard deviation of at least four measurements.\label{LIADrates}}
\end{figure}

Figure~\ref{LIADrates} shows \(\Gamma_{\text{Li}}\) (a), \(K\) (b), and \(P_{f}+P_{s}\)  (c) as a function of LIAD power.
The total asymptotic pressure rise, \(P_{f}+P_{s}\), increases with power and photon energy.
The maximum observed pressure increase is approximately \(50~\si{\percent}\), which occurs when the \(385~\si{\nano\metre}\) LED stimulates desorption.
We expected the background loss coefficient, \(K\), to be independent of power because typical vacuum contaminants have similar collisional properties with Li and the composition of gases adsorbed to the viewport should be similar to the background vapor composition~\cite{Scherschligt2017}.
The data for the \(385~\si{\nano\metre}\) light source are compatible with this expectation.
\(K\) exhibits some variation with power for the two longer wavelength light sources, which may indicate that the gas composition is changing with power.
The spread of the data is large enough that we cannot exclude the possibility of constant \(K\).
The measurements of \(\Gamma_{\text{Li}}\) only show a clear increase with LIAD power for the \(385~\si{\nano\metre}\) source.
The Li-induced loss rate for the \(445~\si{\nano\metre}\) source is consistent with zero at all powers.
This observation supports the presence of a LIAD threshold wavelength \(\gtrsim445~\si{\nano\metre}\), since we expect the Li vapor pressure, and therefore \(\Gamma_{\text{Li}}\), to approach zero near threshold.
In principle, we could use \(\Gamma_{\text{Li}}\) to compute the Li vapor density at the MOT~\cite{Arpornthip2012a, Scherschligt2017}.
However, to do so requires knowledge of kinetic energy distribution of the desorbed lithium atoms.
Because lithium has negligible vapor pressure at room temperature, the untrapped lithium atoms are almost certainly not in thermal equilibrium with the vacuum chamber walls.
Without a more detailed understanding of LIAD, which is beyond the scope of this work, a reasonable calculation of the untrapped Li density is not possible.

\section{\label{Sec:Discussion}Discussion}

We have observed light-induced atomic desorption of lithium and used the desorbed atoms to load a MOT.\@
The MOT contains as many as approximately \(4\times10^{4}\)~\(^{7}\text{Li}\)~atoms and loads at rates as high as approximately \(4\times10^{3}\) atoms per second.
These loading rates are lower than those achieved with lithium dispensers, which can directly load MOTs at rates exceeding \(10^{6}\) atoms per second~\cite{Ladouceur2009, Norrgard2018}.
However, dispenser sources also add a significant gas load to the vacuum system, which could limit the useful lifetime of a compact device~\cite{Rushton2014} or, in the case of vacuum sensing, cause significant systematic effects~\cite{Scherschligt2017}.
By contrast, our LIAD source only increases the pressure during MOT loading by \(\lesssim50~\si{\percent}\).
A LIAD atom source may therefore be preferable for compact vacuum sensors, where a pristine vacuum environment is of greater importance than high atom number.

When assessing the viability of LIAD as an atom source for mobile sensors, we must consider both the maximum achievable atom number and the loading rate.
In general, increasing atom number increases the signal-to-noise ratio.
Increasing the loading rate increases the experimental repetition rate (and measurement duty cycle), which lead to faster averaging, better systematic rejection, and lower aliasing noise~\cite{Meunier2014, Schioppo2017}.
However, how large atom numbers and fast loading rates translate into the measurement of a particular quantity is quite dependent on the details of the measurement itself.
Consider, for example, a single-shot atom interferometry experiment.
In this case, the expected Allan deviation in the phase scales roughly as \(\sqrt{2\tau_{e}/N}\), where \(\tau_{e}\) is the repetition time and \(N\) is the number of atoms.
For the largest MOTs achieved in this work~\cite{Note1}, we might expect an atom shot-noise-limited Allan deviation of the order of \(30~\si{\milli\radian\hertz\tothe{-1/2}}\) (we do not expect that this limit could be realized in a deployable device).
By comparison, a recent mobile interferometer achieved an Allan deviation on the order of \(10~\si{\milli\radian\hertz\tothe{-1/2}}\)~\cite{Hauth2013}.

Vacuum sensors operating below \(10^{-7}~\si{\pascal}\) with an accuracy better than \(10~\si{\percent}\) are not currently available and would be satisfactory for most applications~\cite{Scherschligt2017, Scherschligt2018}.
The relative uncertainty in the lifetime, \(\tau\), of a trapped sample initially containing \(N_{0}\) atoms approaches \(1/\sqrt{N_{0}}\) when measuring the remaining atom number at times around \(2\tau\). 
Thus, in a single shot with \(N_{0}\) known and \(N_{0}\approx 10^{4}\), a sensor should have the necessary precision to exceed the \(10~\si{\percent}\) specification above.
Further averaging is necessary to determine \(N_{0}\) and to look for systematics, but we note that \(\tau_{e}>2\tau\). 
For a \(10^{-8}~\si{\pascal}\) background gas pressure, \(\tau\) is typically of the order of \(10~\si{\second}\), implying that the loading rates we have achieved using the \(385~\si{\nano\metre}\) LED (approximately \(4\times10^{3}\) atoms per second) will not significantly impact this measurement. 

Figure~\ref{LIADpower} suggests two approaches to boosting the MOT atom number and loading rate: increasing the LIAD power or decreasing the LIAD wavelength.
In the limit of negligible background gas pressure, the MOT atom number should saturate to a value given by \(N_{\text{max}}=R/\Gamma_{\text{Li}}\)~\cite{Telles2010, Torralbo-Campo2015} (here we assume that \(R\) and \(\Gamma_{\text{Li}}\) are both directly proportional to the LIAD power).
The data taken using \(385~\si{\nano\metre}\) LED at its highest power output implies \(N_{\text{max}}\approx 10^{5}\).
This analysis suggests that increasing the LIAD power, for the range of wavelengths that we have investigated, will increase the saturated MOT atom number, but only to \(N_{S}\approx N_{\text{max}}\).
Although the potential gain in atom number is limited, using more LIAD power will still increase \(R\).
Our study of the loading rate as a function of the desorption wavelength indicates that the MOT atom number and loading rate could be increased by inducing desorption with a deeper UV light source.
Such a light source would also desorb background gases with higher efficiency, so improvements to the vacuum environment will be necessary to fully realize the potential gain in MOT performance.

Prior studies with other alkalis suggest that LIAD from borosilicate glass, rather than fused silica, yields higher alkali vapor pressures~\cite{Du2004}.
Lithium corrodes most silicate glasses~\cite{Barsoum1981, Maschhoff1986, Bunker1987, Maschhoff1989, Sun2008}, but, to our knowledge, the degradation of vacuum viewports subject to lithium exposure has not been systematically investigated.
The amount of corrosion will presumably be limited due to the small amount of Li deposition necessary for LIAD, but lithium corrosion could shorten the lifetime of compact laser-cooled Li devices.
To circumvent this potential issue, lithium could be deposited on a glass piece contained within the vacuum system~\cite{Du2004} or, possibly, it could be desorbed directly from a pellet of lithium metal~\cite{Kock2016}.
Both of these strategies will be the subject of future experiments.

The authors thank M. Lu and Z. Ahmed, for their careful reading of the manuscript, as well as B. Reschovsky, N. Pisenti, H. Miyake and G. Campbell, for the loan of the blue laser diodes used in this work.
C. Kasik performed several early measurements that contributed to our understanding of LIAD and informed the construction of our apparatus.
DSB acknowledges support from the National Research Council Postdoctoral Research Associateship Program.

\bibliography{LIAD}

\end{document}